\documentclass{article}
\usepackage[T1]{fontenc}
\usepackage[utf8]{inputenc}
\usepackage{ismir,amsmath,cite,url}
\usepackage{graphicx}
\usepackage{color}
\usepackage{multirow}
\usepackage{multicol}
\usepackage{xcolor}
\usepackage{amssymb}
\usepackage{lineno}

\title{Towards robust music source separation \\on loud commercial music}

\oneauthor
  {Chang-Bin Jeon, and Kyogu Lee}
{Department of Intelligence and Information \\ Music and Audio Research Group (MARG) \\ Seoul National University \\ { \tt \{vinyne, kglee\}@snu.ac.kr}}

\def\authorname{C.-B. Jeon, and K. Lee}

\begin{document}

\maketitle
\begin{abstract}
Nowadays, commercial music has extreme loudness and heavily compressed dynamic range compared to the past. Yet, in music source separation, these characteristics have not been thoroughly considered, resulting in the domain mismatch between the laboratory and the real world.
In this paper, we confirmed that this domain mismatch negatively affect the performance of the music source separation networks.
To this end, we first created the out-of-domain evaluation datasets, \textit{musdb-L} and \textit{XL}, by mimicking the music mastering process.
Then, we quantitatively verify that the performance of the state-of-the-art algorithms significantly deteriorated in our datasets. Lastly, we proposed \textit{LimitAug} data augmentation method to reduce the domain mismatch, which utilizes an online limiter during the training data sampling process. We confirmed that it not only alleviates the performance degradation on our out-of-domain datasets, but also results in higher performance on in-domain data.
\end{abstract}

\section{Introduction}\label{sec:introduction}

Recent commercial music has extreme loudness compared to the past \cite{dredge2013pop, milner2019they}. Since many artists and producers want their music to be perceptually louder, it has become so trendy that it rather harms the quality of music \cite{croghan2012quality} and even there exists the expression `loudness war' \cite{vickers2010loudness}.

A dynamic range compressor \cite{stikvoort1986digital} is used to increase the loudness of music while keeping the digital level under 0 decibels relative to full scale (dBFS), which is the maximum possible level in the digital domain.
It is a time-varying non-linear processor that adjusts the level of the signal when the level exceeds the threshold. Especially, a limiter refers to a dynamic range compressor that strongly compresses the signal with a high ratio parameter above 1:10 and increases the gain of the signal by the headroom obtained through compression.

\begin{figure}[t]
  \centering
  \includegraphics[width=5.75cm]{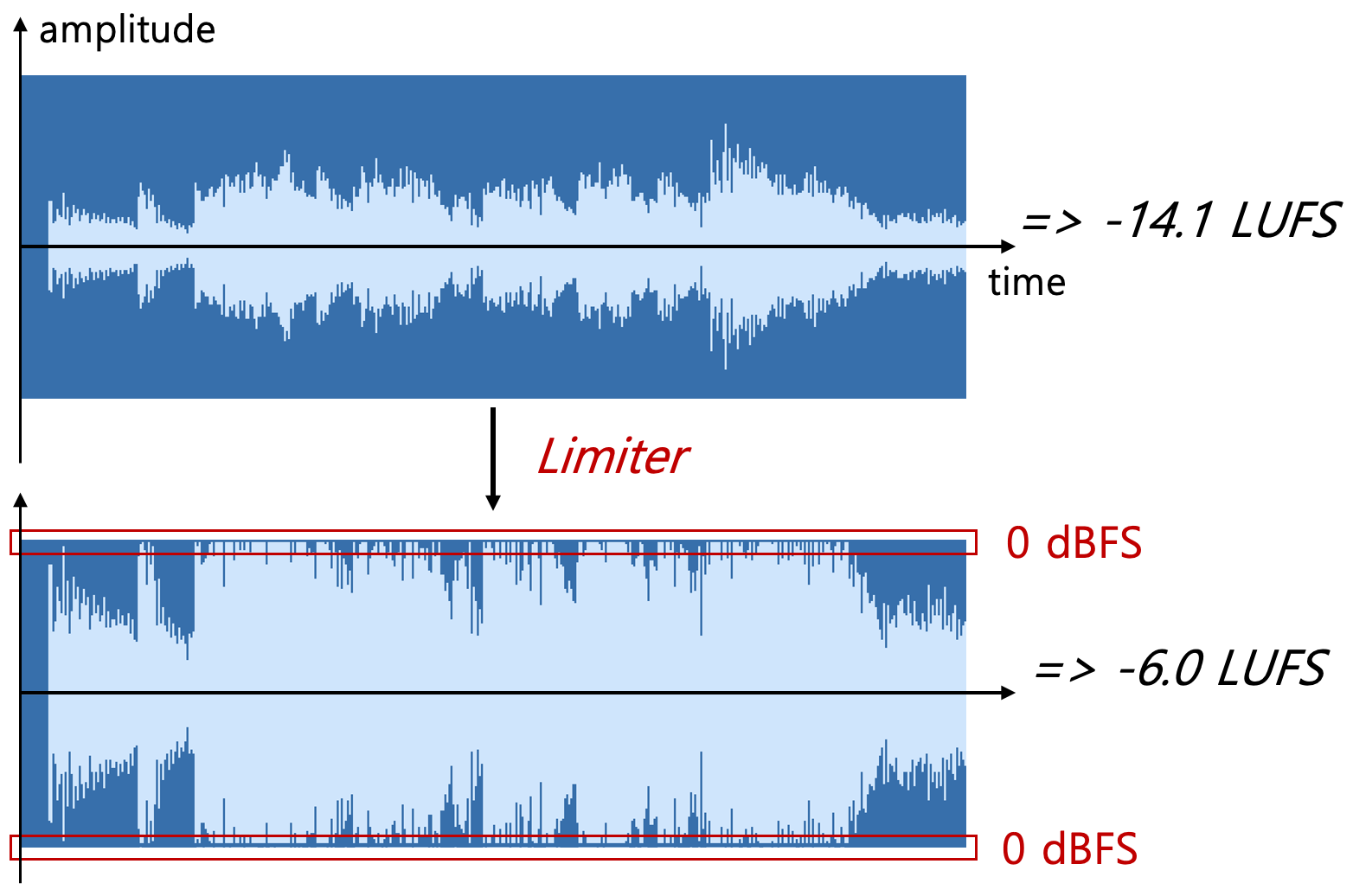}
\vspace{-8pt}
  \caption{The short example of a limiter applied music source in our \textit{musdb-XL} dataset. Recent commercial music has this loud volume and distorted signal characteristics.}
  \label{fig:limiter}
\end{figure}

\begin{figure}[t]
  \centering
  \includegraphics[width=8cm]{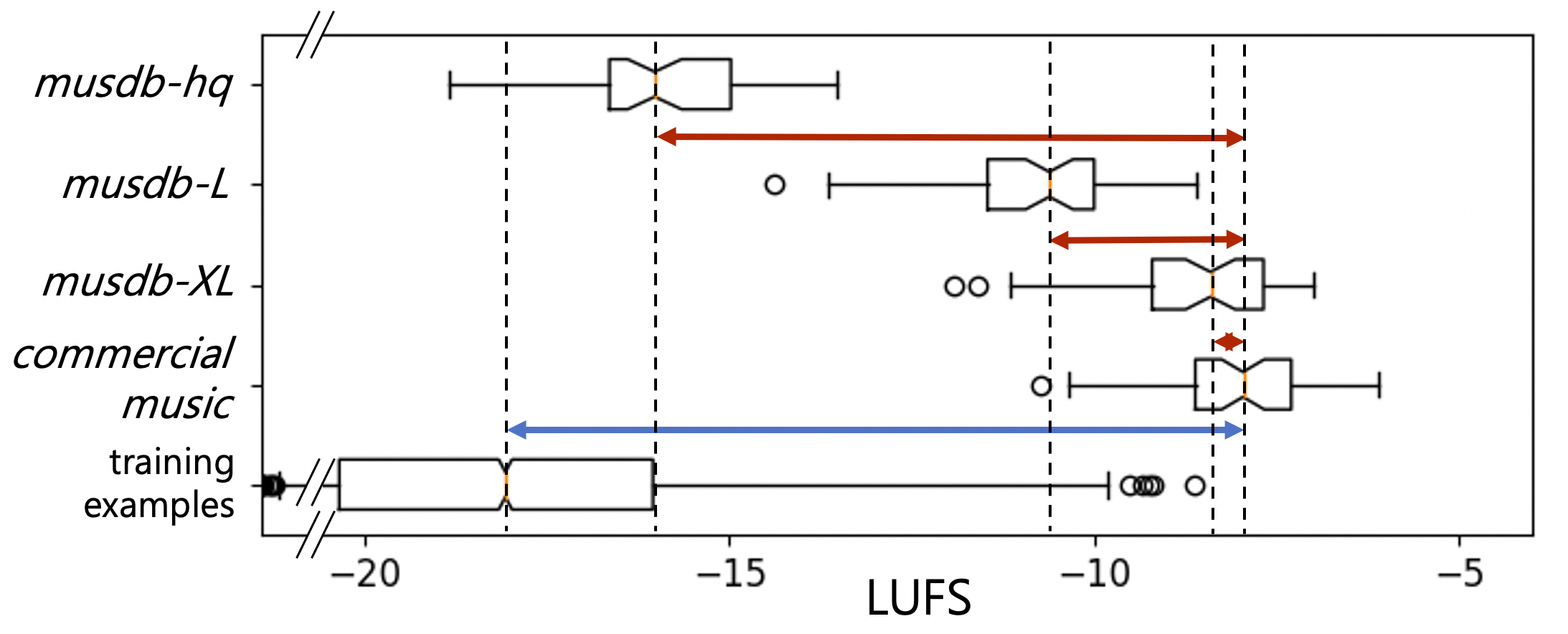}
  \vspace{-8pt}
  \caption{The boxplot representing the loudness distributions of different evaluation datasets, recent commercial music, and training examples generated from the official implementation of \textit{Open-unmix} \cite{stoter19}.}
  \label{fig:domain_shift}
\vspace{-8pt}
\end{figure}

In the last stage of music production, so-called mastering, engineers are often asked to increase the overall loudness of music. A limiter is a core tool for achieving it; it is used to make small parts of music to be louder and loud parts of music fit under 0 dBFS \cite{katz2003mastering, bregitzer2008secrets}. In this process, original signals are distorted and become louder. The example of a limiter usage is depicted in Figure \ref{fig:limiter}.

However, these characteristics have not yet been thoroughly considered in the music source separation. Although many data augmentation techniques have been proposed, such as random gain scaling and mixing of different instrumental stem sources \cite{uhlich2017improving, pretet2019singing}, the training examples have small overall loudness that is far from real-world commercial music\footnote{Based on the investigation of 50 songs in \textit{`Pop Life playlist'} created by Tidal, May. 2022.}, as can be seen in the blue line of Figure \ref{fig:domain_shift}. This causes a huge domain shift between train and real-world domains.
The term LUFS in Figure \ref{fig:domain_shift} is an abbreviate for loudness unit relative to full scale, which is a measure of music loudness \cite{itu2011itu, ebu2011loudness}.

Furthermore, since the standard benchmark datasets for music source separation, \textit{musdb} \cite{musdb18} and \textit{musdb-hq} \cite{musdb18-hq}, do not reflect the characteristics of loud commercial music as shown in the red lines of Figure \ref{fig:domain_shift}, they are unable to show performance degradation comes from the domain shift. Therefore, we conjecture that even good-performing networks based on \textit{musdb} test subset may exhibit worse performance on real world. This problem is no exception to other datasets \cite{manilow2019cutting, liutkus20172016, bittner2014medleydb} for music source separation, since they do not consider the music mastering process.

Based on this question, we investigated how heavily compressed music and its loudness affect the degradation of music source separation networks. To this end, we manually built new evaluation datasets by applying a limiter to the \textit{musdb-hq} test subset \cite{musdb18-hq} to get loud and compressed music imitating the modern music mastering process. One is the \textit{musdb-L} dataset, with increased loudness to an appropriate degree for pleasant sound following \cite{croghan2012quality}. The other is the \textit{musdb-XL} dataset, which raised the loudness to an excessive degree, as often found in recent popular music. Using our datasets, we confirmed that more dynamic range compression in test domains results in more performance degradation of the networks.

Moreover, we conducted various experiments on the training data creating methods to reduce the domain shift between train and real world data from the perspective of music loudness and heavy compression.
We trained music source separation networks using the training examples constructed by \textit{(1)} linear gain increasing, \textit{(2)} the proposed \textit{LimitAug} method which utilizes an online limiter during the sampling process of training examples, \textit{(3)} simple input loudness normalization, and \textit{(4)} the loudness normalization after applying the \textit{LimitAug}.
We confirmed that all of the methods not only showed robust performance on out-of-domain \textit{musdb-L} and \textit{musdb-XL} data, but also showed better performance on in-domain \textit{musdb-hq} test dataset than the baseline.

To summarize, our contributions are three-fold.
\begin{itemize}

\item We built \textit{musdb-L} and \textit{XL} datasets\footnote{https://github.com/jeonchangbin49/musdb-XL}, which have comparable overall loudness to commercial music, for evaluation of music source separation algorithms.
\item Using \textit{musdb-L} and \textit{XL}, we quantitatively confirmed that the domain shift causes performance degradation of the state-of-the-art networks that were trained without considering loud and compressed music characteristics.

\item We proposed \textit{LimitAug}\footnote{https://github.com/jeonchangbin49/LimitAug} data augmentation method and experimentally confirmed that it is beneficial to alleviate the domain shift between train data and the \textit{musdb-L} or \textit{XL}.
\end{itemize}

\section{Musdb-L and Musdb-XL}
\textit{musdb} \cite{musdb18} and \textit{musdb-hq} \cite{musdb18-hq} have been the standard benchmark datasets on music source separation since their presentation in Signal Separation and Evaluation Challenge (SiSEC) 2018 \cite{SiSEC18}. They consist of various genres of 150 professionally produced songs --- 100 songs for train, 50 songs for test --- from folk, indie to electronic, rock genres. Each song has its constituting 4 stems, \textit{vocals}, \textit{drums}, \textit{bass}, and the remaining as \textit{other}. The \textit{musdb-hq} dataset is the uncompressed version of \textit{musdb} with the extended frequency bandwidth from 16kHz to 22.05kHz, which is a full bandwidth of 44.1kHz sample rate. We used \textit{musdb-hq} for high-quality dataset construction.

Since the test subset of \textit{musdb-hq} has small overall loudness compared to commercial mastering-finished music, we conjectured that it could not fully reflect the performance degradation related to heavy dynamic range compression, which prevails in recent commercial music. In addition, to the best of our knowledge, there is no dataset for music source separation that reflects these characteristics or considers the music mastering process. Therefore, we built \textit{musdb-L} and \textit{musdb-XL} datasets, which have loud and compressed characteristics similar to commercial music. \textit{L} and \textit{XL} each stands for \textit{Loud} and \textit{eXtremely Loud}.

\subsection{Dataset construction}
To reflect the characteristics of commercial music, we created datasets by imitating the music mastering process. Both datasets were made by manually applying the commercial digital limiter, iZotope Ozone 9 Maximizer\footnote{https://www.izotope.com/en/products/ozone.html}, to the \textit{musdb-hq} test subset. For each \textit{musdb-L} and \textit{musdb-XL}, we controlled the threshold parameter of a limiter so that compression is applied about 3-4 dB and 6-7 dB in loud parts of \textit{mixture} tracks of the \textit{musdb-hq}. As shown in Table \ref{tab:dataset_loudness}, \textit{musdb-L} and \textit{musdb-XL} are about 5 and 7 LUFS louder than the original \textit{musdb-hq} dataset on average, respectively. \textit{musdb-L} has insufficient loudness compared to loud commercial music but less distorted sources. \textit{musdb-XL} has comparable loudness to commercial music and more distorted sound than \textit{musdb-L}.

To make the ground truth stem tracks of each mixture, we calculated the sample-wise ratio between the limiter applied mixture and the original mixture, then multiply it to the individual stems to make the ground truth stems for \textit{musdb-L} and \textit{XL}.

\begin{table}[t]
\centering

\resizebox{8cm}{!}{
\begin{tabular}{ccccc}

\hline
\multirow{2}{*}{\textbf{dataset}} & \multicolumn{4}{c}{\textbf{Loudness {[}LUFS{]}}}                                                         \\ \cline{2-5} 
                         & min         & max    & median         & mean (std)                          \\ \hline
\textit{musdb-hq}      & \multicolumn{1}{l}{-18.84} & \multicolumn{1}{l}{-13.52} & \multicolumn{1}{l}{-16.02} & \multicolumn{1}{l}{-15.92 (1.27)} \\
\textit{musdb-L}      & -14.39                     & -8.61 & -10.61      & -10.89 (1.19)                     \\
\textit{musdb-XL} & -11.93                     & -6.99  & -8.41                   & -8.61 (1.17)                      \\ 
\hline\hline
\textit{commercial} & \textbf{-10.75}                     & \textbf{-6.10}  &-\textbf{7.96}                    & \textbf{-8.05 (1.06)}                      \\ \hline 
\end{tabular}
}
\caption{Loudness statistics of \textit{musdb-hq}, \textit{musdb-L}, \textit{musdb-XL} datasets, and the investigated commercial music.}
\vspace{-8pt}
\label{tab:dataset_loudness}
\end{table}

\section{Methods}\label{sec:methods}
In general, a limiter is used as the last signal processor in music mastering of commercial music \cite{katz2003mastering, bregitzer2008secrets}. Since it tends to be used excessively in modern commercial music \cite{dredge2013pop, milner2019they, vickers2010loudness}, it should be considered in music source separation for the development of robust application.

We assumed that the key differences between the real-world mastering-finished music, i.e. a limiter applied music, and the standard training examples for music source separation are \textit{\romannumeral 1)} the overall amplitude scale, and \textit{\romannumeral 2)} the signal distortion caused by a limiter.

Here we introduce two ways to avoid each domain mismatch problem.

\subsection{Input loudness normalization}
First of all, the simple loudness normalization, which is a linear gain adjustment of network inputs to the pre-defined reference level, is the easiest technique to avoid the domain shift caused by \textit{\romannumeral 1)} overall amplitude scale mismatch. This can be categorized into two, \textit{(\romannumeral 1)} input loudness normalization during both the training and evaluation stages of networks, and \textit{(\romannumeral 2)} normalization only at the evaluation stage. 

The method \textit{(\romannumeral 1)} is already used in various studies with different ways. For example, the network such as \textit{Demucs v3} \cite{defossez2021hybrid}, uses input standardization in time domain based on the mean and standard deviation of the training data. In \textit{Open-unmix} \cite{stoter19}, the network implicitly normalizes the input by utilizing trainable input scaling parameters that works in time-frequency domain. However, we assumed that explicitly adjusting the gain of the inputs based on the waveform, thereby minimizing the overall amplitude scale mismatch between train and test domain, can greatly reduce the performance degradation comes from the domain shift.

The method \textit{(\romannumeral 2)}, normalization only at the evaluation stage, can be considered as a readily applicable workaround for the models that were trained without considering music loudness. This method is a simple idea to reduce the overall amplitude scale difference between train and real-world domain. Since the music source separation networks are non-linear systems, we hypothesized that linear scale difference of the inputs might affect the quality of final outputs. That is, input normalization only at the evaluation stage can be a simple, yet effective trick for models that were already trained without considering the domain mismatch.

Though these methods can mitigate the domain mismatch related to  \textit{\romannumeral 1)} amplitude scale, there needs to be another strategies that can reflect \textit{\romannumeral 2)} the signal distortion caused by a limiter.

\subsection{\textit{LimitAug}}
The best way to consider the characteristics of the limiter applied source is simply using the limiter during the network training. Therefore, we propose the \textit{LimitAug} data augmentation method, which utilizes an online limiter during the training examples construction process, to reduce the domain shift between the training examples and real-world commercial music. 
The \textit{LimitAug} can be considered as creating train examples that forcibly reflect the distortion that comes from a limiter, which cannot be reflected by the simple input loudness normalization technique.

\begin{figure}[t]\vspace{-5pt}
\centerline{\includegraphics[width=0.78\columnwidth]{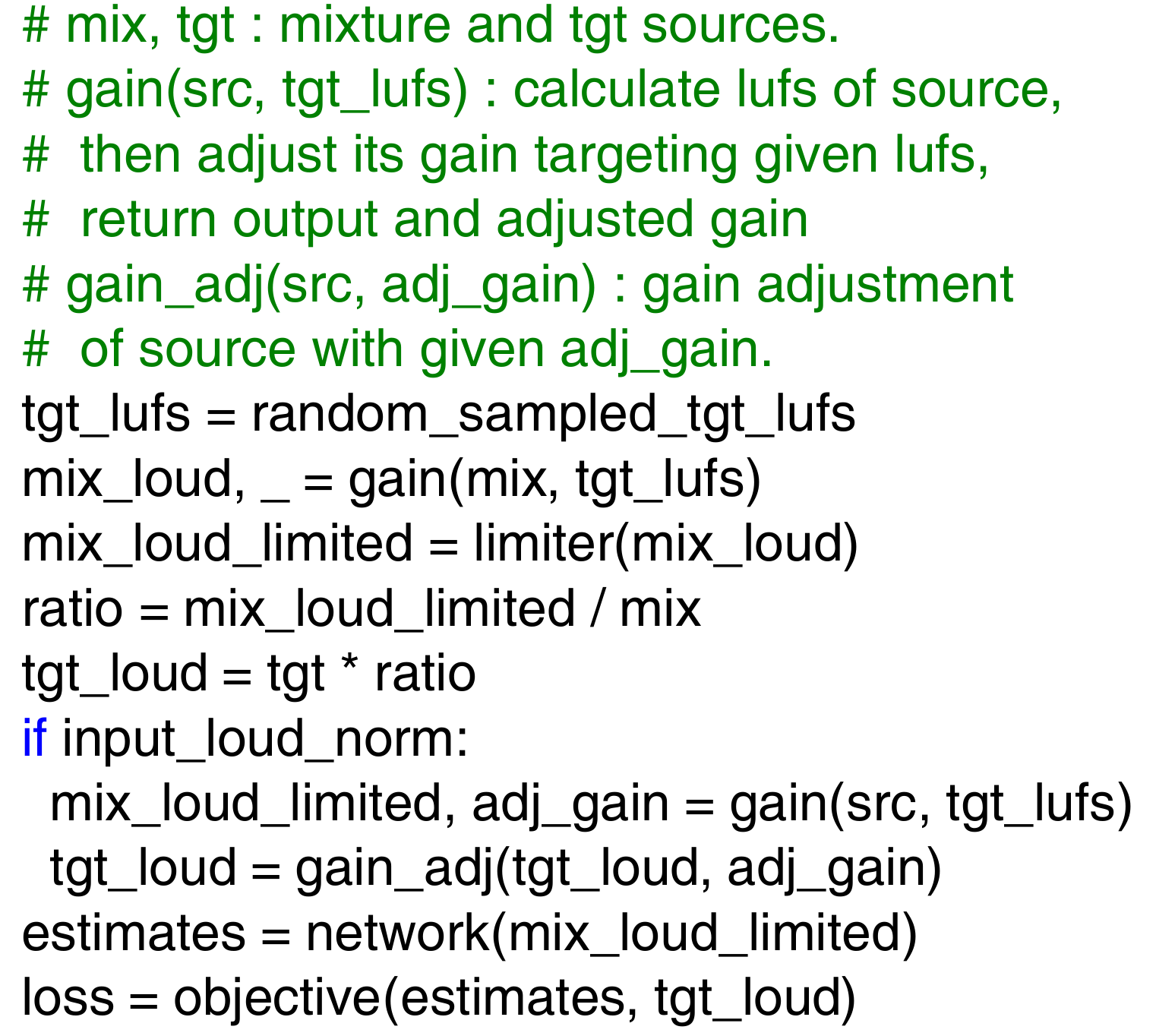}}
\vspace{-10pt}
\caption{\texttt{numpy}-like pseudocode of the proposed \textit{LimitAug} method.}
\vspace{-8pt}
\label{fig:limitaug_pseudocode}
\end{figure}

The pseudocode for \textit{LimitAug} is shown in Figure \ref{fig:limitaug_pseudocode}. First, calculate the LUFS of the mixture created by the data sampling process.
Second, adjust the gain of the input mixture source targeting the randomly sampled LUFS value.
Then, it is followed by the online limiter to fit the waveform under 0 dBFS. Lastly, calculate the sample-wise (A sample refers to an each waveform value) ratio between the limiter applied mixture source and the original mixture source, then multiply the ratio to the original target source to get the ground truth target signal of the limiter applied mixture source.

When adjusting the gain and applying the limiter to the input mixture, for example, if the input mixture had -15 LUFS and the randomly sampled target LUFS was -10, note that the gain-scaled mixture by +5 dB does not have exact -10 LUFS due to the compression of a limiter and the nature of LUFS calculation, which considers frequency weighting \cite{itu2011itu, ebu2011loudness}.

The proposed \textit{LimitAug} can be used with other data augmentation methods; in our study, random gain scaling, channel swap and mixing of different instrumental stem sources \cite{uhlich2017improving, pretet2019singing} were used. Also, additional loudness normalization can be applied after the \textit{LimitAug} as depicted in conditional statement of Figure \ref{fig:limitaug_pseudocode}, so that the overall amplitude scale mismatch between training and evaluation stages be minimized.

\begin{table*}[t]
\centering
\resizebox{15cm}{!}
{
\begin{tabular}{cccccccc}
\hline
\multirow{2}{*}{\textbf{network}}                                         & \multirow{2}{*}{\textbf{\begin{tabular}[c]{@{}c@{}}extra\\ train data\end{tabular}}} & \multirow{2}{*}{\textbf{\begin{tabular}[c]{@{}c@{}}test\\ data\end{tabular}}} & \multicolumn{5}{c}{\textbf{SDR median (mean) {[}dB{]}}}                                                          \\ \cline{4-8} 
                                                                          &                                                                                      &                                                                               & \textit{vocals}               & \textit{bass}                 & \textit{drums}                & \textit{other}                & \textit{avg}                  \\ \hline
\multirow{3}{*}{\textit{Open-unmix} \cite{stoter19}}                                               & \multirow{3}{*}{-}                                                                   & \textit{hq}                                                                            & 6.16 (2.54)          & 5.03 (2.67)          & 6.00 (5.46)          & 4.22 (3.46)          & 5.35 (3.53)          \\
                                                                          &                                                                                      & \textit{L}                                                                             & 6.33 (1.63)          & 4.81 (2.71)          & 5.82 (5.38)          & 4.11 (3.42)          & 5.27 (3.28)          \\
                                                                          &                                                                                      & \textit{XL}                                                                            & 5.98 (0.89)          & 4.76 (2.59)          & 4.97 (4.89)          & 4.04 (3.29)          & 4.94 (2.92)          \\ \hline
\multirow{3}{*}{\begin{tabular}[c]{@{}c@{}}\textit{TFC-TDF}\\ \textit{-U-net} \cite{choi2020investigating}\end{tabular}} & \multirow{3}{*}{-}                                                                   & \textit{hq}                                                                            & 7.18 (4.26)          & 5.59 (3.35)          & 5.76 (5.30)          & 4.04 (3.18)          & 5.64 (4.02)          \\
                                                                          &                                                                                      & \textit{L}                                                                             & 7.03 (3.65)          & 5.41 (3.08)          & 5.52 (5.09)          & 3.67 (3.00)          & 5.41 (3.71)          \\
                                                                          &                                                                                      & \textit{XL}                                                                            & 6.95 (3.14)          & 5.48 (2.90)          & 5.11 (4.68)          & 3.55 (2.82)          & 5.27 (3.39)          \\ \hline

\multirow{3}{*}{\textit{Demucs v3-A} \cite{defossez2021hybrid}}                                            & \multirow{3}{*}{-}                                                                   & \textit{hq}                                                                            & \textbf{8.11} \textbf{(5.22)}          & \textbf{9.34} (6.21)          & \textbf{8.57} \textbf{(8.01)}          & \textbf{5.51} \textbf{(5.03)}          & \textbf{7.88} \textbf{(6.12)}          \\
                                                                          &                                                                                      & \textit{L}                                                                             & 7.54 (5.15)          & 9.32 \textbf{(6.22)}          & 8.26 (7.65)          & 5.51 (5.01)          & 7.66 (6.01)          \\
                                                                          &                                                                                      & \textit{XL}                                                                            & 7.30 (4.86)          & 9.19 (6.14)          & 7.62 (6.78)          & 5.37 (4.97)          & 7.37 (5.69)          \\ \hline\hline
\multirow{3}{*}{\textit{Open-unmix} \cite{stoter19}}                                               & \multirow{3}{*}{\checkmark}                                                                   & \textit{hq}                                                                            & 7.02 (4.93)          & 5.91 (4.06)          & 7.18 (6.91)          & 4.94 (4.76)          & 6.26 (5.17)          \\
                                                                          &                                                                                      & \textit{L}                                                                             & 6.83 (5.12)          & 6.23 (4.09)          & 7.07 (6.92)          & 4.94 (4.78)          & 6.27 (5.23)          \\
                                                                          &                                                                                      & \textit{XL}                                                                            & 6.70 (4.77)          & 6.16 (3.87)          & 6.80 (6.48)          & 4.89 (4.61)          & 6.14 (4.93)          \\ \hline
\multirow{3}{*}{\textit{Spleeter} \cite{hennequin2020spleeter}}                                                 & \multirow{3}{*}{25000+}                                                                   & \textit{hq}                                                                            &   6.51 (4.42)                   &              4.77 (3.57)       &       6.00 (6.09)               &          4.22 (4.12)            &      5.38 (4.55)                \\
                                                                          &                                                                                      & \textit{L}                                                                             &   6.18 (3.90)                   &  4.73 (3.34)                    &    5.67 (5.94)                  &    4.37 (4.03)                  &        5.24 (4.30)              \\
                                                                          &                                                                                      & \textit{XL}                                                                            &  6.03 (3.38)                    &    4.80 (3.13)                  &      5.55 (5.52)                &        4.24 (3.91)              &  5.15 (3.98)                    \\ \hline
\multirow{3}{*}{\textit{Demucs v3-B} \cite{defossez2021hybrid}}                                            & \multirow{3}{*}{\begin{tabular}[c]{@{}c@{}}200+ including\\ \textit{musdb-hq} test set\end{tabular} }                                                                   & \textit{hq}                                                                            & \textbf{9.24} \textbf{(7.05)}          & \textbf{11.65 (9.58)}         & \textbf{11.73 (11.34)}        & \textbf{7.83 (8.03)}          & \textbf{10.11 (9.00)}         \\
                                                                          &                                                                                      & \textit{L}                                                                             & 9.05 (6.91)          & 11.61 (9.55)         & 11.05 (10.27)        & 7.83 (7.91)          & 9.88 (8.66)          \\
                                                                          &                                                                                      & \textit{XL}                                                                            & 8.76 (6.41)          & 11.56 (9.29)         & 9.22 (8.78)          & 7.52 (7.51)          & 9.26 (8.00)          \\ \hline
\end{tabular}
}
\caption{Performance of various music source separation networks trained with \textit{musdb-hq} \cite{musdb18-hq} on the test using \textit{musdb-hq}, \textit{musdb-L} and \textit{musdb-XL}.}
\vspace{-8pt}
\label{tab:3models}
\end{table*}

\section{Experiments}
Here we briefly summarize our following experiments.

In Section \ref{sec:degradation}, we quantitatively evaluated various music source separation networks that were trained without considering the loudness and heavy dynamic range compression, and confirmed the performance degradation on \textit{musdb-L} and \textit{musdb-XL}, compared to the origianl \textit{musdb-hq} evaluation dataset.

In Section \ref{sec:degradation_normalization}, we simply normalized the network inputs in the evaluation stage, to observe the performance degradation caused by the signal distortion caused by a limiter, not an overall amplitude mismatch between the training and the evaluation data. Then, we carefully analyzed the results on \textit{Demucs v3} \cite{defossez2021hybrid}, which took the 1st place in the 2021 Sony Music Demixing (MDX) Challenge leaderboard A and the 2nd place in the leaderboard B \cite{mitsufuji2021music}. The leaderboard A was the competition that only allowed training on the only \textit{musdb-hq} train subset, and the leaderboard B allowed the extra training data.

In Section \ref{sec:various_strategies}, we conducted a comparative study on various training data construction strategies. Unfortunately, we were unable to train the current state-of-the-art \textit{Demucs v3}, due to the constraint of our GPU experimental environments.
Considering the training efficiency and reasonable performance, we used \textit{TFC-TDF-U-Net} \cite{choi2020investigating} --- a backbone architecture of KUIELab-MDX-Net \cite{kim2021kuielab} with slight modifications, which took the 2nd place in the 2021 MDX Challenge leaderboard A --- in this experiment.

\subsection{Training}
In Section \ref{sec:degradation} and Table \ref{tab:3models}, we used official pre-trained weights of each model except \textit{TFC-TDF-U-Net} since there were no official weights for the 4 stems of \textit{musdb-hq}. For fair comparison in Section \ref{sec:various_strategies}, we trained \textit{TFC-TDF-U-Net} for 300 epochs with early stopping, based on official training framework of \textit{Open-unmix} \cite{stoter19} and official network implementation of \textit{TFC-TDF-U-net}. We used default network hyper-parameters introduced in its webpage\footnote{https://github.com/ws-choi/ISMIR2020\_U\_Nets\_SVS}.

For the \textit{LimitAug}, we used the limiter implemented in \texttt{pedalboard} \cite{pedalboard}. The threshold parameter was set to 0 dBFS, and the release parameter was randomly sampled from uniform distribution of (30, 200) millisecond in data sampling process. Also, we used \texttt{pyloudnorm} \cite{steinmetz2021pyloudnorm} for loudness calculation.

\subsection{Evaluation}
In all of the evaluations, Signal-to-Distortion Ratio (SDR) \cite{vincent2006performance} was calculated using \texttt{museval} python library \cite{SiSEC18}. Also, in following experimental results, both \textit{median} and \textit{mean} SDR scores were presented for the detailed comparison. Note that we only used the \textit{musdb-hq} train subset for training the networks. \textit{musdb-L} and \textit{XL} are only for evaluation. 

In the original \textit{musdb-hq} test subset, as stated in the official webpage\footnote{https://sigsep.github.io/datasets/musdb}, `PR - Oh No' track's mixture signal is panned to the right channel, which causes the inconsistency between linear summation of stems and mixture source. Since this causes the limiter to be operated in un-musical way while the construction of \textit{musdb-L} and \textit{XL}, which also can be hardly found in popular music, we used the linear summation of stems as a mixture only for this track. This may results in the slight difference of SDR scores between the Table \ref{tab:3models} and the official scores of each network.

\begin{table*}[t]
\centering
\resizebox{14.5cm}{!}
{
\begin{tabular}{cccccccc}
\hline
\multirow{2}{*}{\textbf{network}}                                         & \multirow{2}{*}{\textbf{\begin{tabular}[c]{@{}c@{}}extra\\ train data\end{tabular}}} & \multirow{2}{*}{\textbf{\begin{tabular}[c]{@{}c@{}}test\\ data\end{tabular}}} & \multicolumn{5}{c}{\textbf{SDR median (mean) {[}dB{]}}}                                                          \\ \cline{4-8} 
                                                                          &                                                                                      &                                                                               & \textit{vocals}               & \textit{bass}                 & \textit{drums}                & \textit{other}                & \textit{avg}                  \\ \hline
\multirow{3}{*}{\textit{Demucs v3-A} \cite{defossez2021hybrid}}                                            & \multirow{3}{*}{-}                                                                   & \textit{hq}                                                                            & \textbf{8.11} (5.22)          & \textbf{9.34 (6.21)}          & \textbf{8.57 (8.01)}          & 5.51 \textbf{(5.03)}          & \textbf{7.88 (6.12)}          \\
                                                                          &                                                                                      & \textit{L}                                                                             & 8.05 \textbf{(5.23)}          & 9.25 (6.20)          & 8.47 (7.92)          & \textbf{5.53} (5.02)          & 7.82 (6.09)          \\
                                                                          &                                                                                      & \textit{XL}                                                                            & 7.93 (5.03)          & 9.27 (5.92)          & 7.74 (7.44)          & 5.55 (4.91)          & 7.62 (5.82)          \\ \hline\hline
\multirow{3}{*}{\textit{Demucs v3-B} \cite{defossez2021hybrid}}                                            & \multirow{3}{*}{\begin{tabular}[c]{@{}c@{}}200+ including\\ \textit{musdb-hq} test set\end{tabular}}                                                                   & \textit{hq}                                                                            & \textbf{9.24 (7.05)}          & \textbf{11.65 (9.58)}         & \textbf{11.73 (11.34)}        & \textbf{7.83 (8.03)}          & \textbf{10.11 (9.00)}         \\
                                                                          &                                                                                      & \textit{L}                                                                             & 9.19 (7.04)          & 11.64 (9.55)         & 11.68 (11.21)        & 7.82 (8.02)          & 10.08 (8.95)          \\
                                                                          &                                                                                      & \textit{XL}                                                                            & 9.13 (6.90)          & 11.56 (9.33)         & 11.32 (10.75)          & 7.74 (7.95)          & 9.94 (8.73)          \\ \hline
\end{tabular}
}
\caption{Performance of \textit{Demucs v3} \cite{defossez2021hybrid} trained with \textit{musdb-hq} \cite{musdb18-hq}. On the test using \textit{musdb-L} and \textit{musdb-XL}, each of the input sources were loudness normalized targeting the LUFS of its original \textit{musdb-hq} sources.}
\vspace{-6pt}
\label{tab:demucs_stems}
\end{table*}

\begin{table}[t]
\centering
\resizebox{8cm}{!}{
\begin{tabular}{ccccc}
\hline
\multirow{2}{*}{\textbf{network}} & \multirow{2}{*}{\begin{tabular}[c]{@{}c@{}}\textbf{extra}\\ \textbf{train data}\end{tabular}} & \multicolumn{3}{c}{\textbf{SDR median {[}dB{]}}} \\ \cline{3-5} 
                         &                                                                             & \textit{hq}           & \textit{L}           & \textit{XL}         \\ \hline
\textit{Open-unmix} \cite{stoter19}               & -                                                                           & 5.35         & 5.32        & 5.25       \\ 
\textit{TFC-TDF-U-Net} \cite{choi2020investigating}            & -                                                                           & 5.64         & 5.62        & 5.51       \\ 
\textit{Demucs v3-A} \cite{defossez2021hybrid}           & -                                                                           & \textbf{7.88}         & \textbf{7.82}        & \textbf{7.62}       \\ \hline\hline
\textit{Open-unmix} \cite{stoter19}              & \checkmark                                                                           & 6.26         & 6.25     &   6.18         \\ 
\textit{Spleeter} \cite{hennequin2020spleeter}                 & \checkmark                                                                           & 5.38         &  5.33           &            5.21\\ 
\textit{Demucs v3-B} \cite{defossez2021hybrid}           & \checkmark                                                                           & \textbf{10.11}       &  \textbf{10.08}      &    \textbf{9.94}       \\ \hline
\end{tabular}
}
\caption{Performance of networks with loudness normalized input at the evaluation stage. Each score represents the average score across the 4 stems. Note that the networks were trained without considering music loudness or dynamic range compression explicitly.}
\vspace{-10pt}
\label{tab:loudnorm_eval}
\end{table}

\section{Results} \label{sec:results}

\subsection{Performance degradation on \textit{musdb-L} and \textit{XL}} \label{sec:degradation}

Here we quantitatively evaluated the performance of state-of-the-art networks on \textit{musdb-hq} \cite{musdb18-hq}, \textit{L} and \textit{XL} datasets and confirmed that the domain shift in perspective of music loudness and compression negatively affect the performance, indeed. As shown in the Table \ref{tab:3models}, all networks showed significant performance degradation on the evaluations with \textit{musdb-L} and \textit{musdb-XL} datasets. 
The amount of decrease on \textit{Demucs v3} \cite{defossez2021hybrid} was slightly larger than the others. Overall, we concluded that every networks are somewhat overfitted to the \textit{musdb-hq} data, making the networks vulnerable to loud and heavily compressed music. Therefore, it is highly required to consider these characteristics for the robust music source separation.
Note that we stated the scores on different test datasets in each block to emphasize the performance degradation between the domains.

\subsubsection{Extra training data}
Though the extra training data was of help, it did not work as the fundamental solution to the domain shift. \textit{Open-unmix} \cite{stoter19} with extra training data showed more robust performance than that of the network without extra data, but performance degradation on \textit{Demucs v3-B} was more significant than that of \textit{Demucs v3-A}. Since \textit{Demucs v3-B} was trained with extra training data including \textit{musdb-hq} test subset, it is reasonable to guess that the network is highly overfitted to the \textit{musdb-hq} data. Nevertheless, the amount of performance decrease especially on \textit{drums} comparing the scores between \textit{musdb-hq} and \textit{XL}, 2.5 dB, is somewhat high. Considering \textit{Demucs v3} uses an input standardization technique based on mean and standard deviation of waveform values both at training and evaluation stages, this implies that there needs to be another solution for the robust music source separation.

\subsubsection{Performance degradation on \textit{drums} and \textit{vocals}}\label{sec:drums_and_vocals}

It is noteworthy that the degradation on \textit{drums} and \textit{vocals} were more significant than the others. Due to the percussive nature of \textit{drums}, in general, they have the biggest momentary energy in music. Also, since \textit{vocals} are important ingredients in modern commercial music, they usually consist of not only a single singing voice but also doubling, harmony and chorus. Therefore, we assumed \textit{drums} and \textit{vocals} are most affected by the limiter, which is activated when loud input sources that are above the threshold are given. To quantitatively confirm the signal distortion by the limiter, we calculated Scale-invariant Signal-to-Distortion Ratio (SI-SDR) \cite{le2019sdr} between \textit{musdb-hq} and \textit{musdb-XL} for each stem. As a result, \textit{drums} and \textit{vocals} scored each 19.97 and 23.69 dB, on the other hand, \textit{bass} and \textit{other} scored each 25.12 and 25.48 dB on average. That is, the signal distortion caused by a limiter is more significant on \textit{drums} and \textit{vocals}. 

Unfortunately, since the networks in the Table \ref{tab:3models} have never seen these kinds of distorted \textit{drums} or \textit{vocals} as training examples, we assumed that the degradation on \textit{drums} and \textit{vocals} were significant compared to the rest. This result strongly supports the necessity of considering the heavy dynamic range compression from the training stage, i.e. the \textit{LimitAug}, which forcibly makes the distorted and compressed training examples for training purposes, thereby minimizing the domain shift.

\begin{table*}[t]
\centering
\resizebox{17cm}{!}
{
\begin{tabular}{cccccccccc}
\hline
\multirow{2}{*}{\textbf{network}} & \multirow{2}{*}{\textbf{methods}} & \multirow{2}{*}{\begin{tabular}[c]{@{}c@{}}\textbf{linear} \\ \textbf{gain increase}\end{tabular}} & \multirow{2}{*}{\textit{\textbf{LimitAug}}} & \multirow{2}{*}{\begin{tabular}[c]{@{}c@{}}\textbf{input}\\ \textbf{loud-norm}\end{tabular}} & \multirow{2}{*}{\textbf{target LUFS}} & \multicolumn{4}{c}{\textbf{SDR median (mean) {[}dB{]}}} \\ \cline{7-10} 
                                  &                                   &                                                                                  &                           &                                                                            &                              & \textit{hq}           & \textit{L}            & \textit{XL}          & \textit{avg}         \\ \hline
\multirow{8}{*}{\textit{\begin{tabular}[c]{@{}c@{}}TFC-TDF\\ -U-Net \cite{choi2020investigating}\end{tabular}}}    & baseline                          & -                                                                                & -                         & -                                                                          & -                            & 5.64 (4.02)  & 5.41 (3.71)  & 5.27 (3.39) & 5.44 (3.71) \\ \cline{2-10} 
                                  & \multirow{2}{*}{\textit{(1)}}              & \multirow{2}{*}{\checkmark}                                                               & \multirow{2}{*}{-}        & \multirow{2}{*}{-}                                                         & \textit{$\mathcal{N}(\mu_{L},\sigma_{L}^{2})$}                         & \textbf{5.90} (4.31)  & 5.86 (4.33)  & 5.73 (4.15) & 5.83 (4.26) \\ 
                                  &                                   &                                                                                  &                           &                                                                            & \textit{$\mathcal{N}(\mu_{XL},\sigma_{XL}^{2})$}                        & 5.32 (3.43)  & 5.36 (3.62)  & 5.28 (3.49) & 5.32 (3.51) \\ \cline{2-10} 
                                  
                                  & \multirow{2}{*}{\textit{(2)}}              & \multirow{2}{*}{-}                                                               & \multirow{2}{*}{\checkmark}        & \multirow{2}{*}{-}                                                         & \textit{$\mathcal{N}(\mu_{L},\sigma_{L}^{2})$}                         & 5.79 (4.30)  & \textbf{5.90} \textbf{(4.41)}  & 5.74 (4.25) & 5.81 (4.32) \\  
                                  &                                   &                                                                                  &                           &                                                                            & \textit{$\mathcal{N}(\mu_{XL},\sigma_{XL}^{2})$}                        & 5.69 (3.93)  & 5.72 (4.22)  & 5.57 (4.15) & 5.66 (4.10) \\ \cline{2-10} 
                                  & \textit{(3)}                               & -                                                                                & -                         & \checkmark                                                                          & -14                          & 5.89 \textbf{(4.38)}  & 5.87 (4.35)  & \textbf{5.82 (4.25)} & \textbf{5.86 (4.33)} \\ \cline{2-10} 
                                  & \multirow{2}{*}{\textit{(4)}}              & \multirow{2}{*}{-}                                                               & \multirow{2}{*}{\checkmark}        & \multirow{2}{*}{\checkmark}                                                         & \textit{$\mathcal{N}(\mu_{L},\sigma_{L}^{2})$}, -14                         & 5.87 (4.25)  & 5.85 (4.21)  & 5.76 (4.16) & 5.83 (4.21) \\  
                                  &                                   &                                                                                  &                           &                                                                            & \textit{$\mathcal{N}(\mu_{XL},\sigma_{XL}^{2})$}, -14                        & 5.78 (4.27)  & 5.78 (4.26)  & 5.73 (4.20) & 5.76 (4.24) \\ \hline
\end{tabular}
}
\caption{Performance of \textit{TFC-TDF-U-Net} \cite{choi2020investigating} trained with various training data construction strategies. Each score represents the average score across the 4 stems.}
\label{tab:training_strategies}
\end{table*}

\begin{table*}[t]
\centering
\resizebox{15cm}{!}
{
\begin{tabular}{ccccccccc}
\hline
\multirow{2}{*}{\textbf{network}}& \multirow{2}{*}{\textbf{methods}}&\multirow{2}{*}{\begin{tabular}[c]{@{}c@{}}\textbf{target}\\ \textbf{LUFS}\end{tabular}} & \multirow{2}{*}{\begin{tabular}[c]{@{}c@{}}\textbf{test}\\ \textbf{data}\end{tabular}} & \multicolumn{5}{c}{\textbf{SDR median (mean) {[}dB{]}}}                      \\ \cline{5-9} 
                              &                                                                                     &                                                                        &                                                                      & \textit{vocals}      & \textit{bass}        & \textit{drums}       & \textit{other}       & \textit{avg}         \\ \hline
\multirow{9}{*}{\textit{\begin{tabular}[c]{@{}c@{}}TFC-TDF\\ -U-Net \cite{choi2020investigating}\end{tabular}}} & \multirow{3}{*}{baseline}                                                           & \multirow{3}{*}{-}                                                     & \textit{hq}                                                                   & 7.18 (4.26) & 5.59 (3.35) & 5.76 (5.30) & 4.04 (3.18) & 5.64 (4.02) \\  
                              &                                                                                     &                                                                        & \textit{L}                                                                    & 7.03 (3.65) & 5.41 (3.08) & 5.52 (5.09) & 3.67 (3.00) & 5.41 (3.71) \\  
                              &                                                                                     &                                                                        & \textit{XL}                                                                   & 6.95 (3.14) & 5.48 (2.90) & 5.11 (4.68) & 3.55 (2.82) & 5.27 (3.39) \\ \cline{2-9} 
                              & \multirow{3}{*}{\textit{(3)} loud-norm}                                                      & \multirow{3}{*}{-14}                                                   & \textit{hq}                                                                   & 7.35 \textbf{(4.76)} & \textbf{5.93} (3.61) & \textbf{5.91} \textbf{(5.37)} & 4.39 (3.79) & \textbf{5.89} \textbf{(4.38)} \\  
                              &                                                                                     &                                                                        & \textit{L}                                                                    & 7.32 (4.72) & 5.91 (3.61) & 5.85 (5.29) & 4.39 (3.78) & 5.87 (4.35) \\ 
                              &                                                                                     &                                                                        & \textit{XL}                                                                   & 7.26 (4.64) & 5.91 \textbf{(3.62)} & 5.68 (4.99) & 4.42 (3.78) & 5.82 (4.25) \\ \cline{2-9} 
                              & \multirow{3}{*}{\begin{tabular}[c]{@{}c@{}}\textit{(4)} \textit{LimitAug},\\ loud-norm\end{tabular}} & \multirow{3}{*}{\begin{tabular}[c]{@{}c@{}}\textit{$\mathcal{N}(\mu_{L},\sigma_{L}^{2})$,}\\-14 \end{tabular}} & \textit{hq} & \textbf{7.59} (4.64) & 5.75 (3.25) & 5.63 (5.28) & 4.50 (3.82) & 5.87 (4.25) \\  
                              & & & \textit{L} & 7.58 (4.61) & 5.69 (3.21) & 5.62 (5.22) & 4.50 (3.82) & 5.85 (4.21) \\  
                              & & & \textit{XL}& 7.48 (4.55) & 5.67 (3.29) & 5.36 (4.99) & \textbf{4.51} \textbf{(3.82)} &    5.76 (4.16)         \\ \hline
\end{tabular}
}
\caption{Stem-wise performance of \textit{TFC-TDF-U-Net} \cite{choi2020investigating} trained with the method \textit{(3)} input loudness normalization, and \textit{(4)} input loudness normalization after the proposed \textit{LimitAug}.}
\vspace{-6pt}
\label{tab:best_performing}
\end{table*}

\subsection{Analysis on the input normalization at the evaluation stage} \label{sec:degradation_normalization}
If the key differences between the real-world music and the training examples of music source separation networks are \textit{\romannumeral 1)} overall amplitude scale, and \textit{\romannumeral 2)} the signal distortion caused by a limiter, as stated in Section \ref {sec:methods}, then what if we give the loudness normalized \textit{musdb-L} or \textit{XL} data as inputs to the networks in Table \ref{tab:3models}? Due to the non-linear nature of deep neural networks, we assumed that simply normalizing the amplitude scale of the networks on the evaluation stage may affect the performance.

The inference was conducted with following procedures, \textit{(\romannumeral 1)} reducing the loudness of \textit{musdb-L} or \textit{XL} input so that its loudness becomes same with that of the corresponding original \textit{musdb-hq} data, \textit{(\romannumeral 2)} inference with loudness normalized input, and \textit{(\romannumeral 3)} increasing the scale of output as much as reduced in step \textit{(\romannumeral 1)}.

Comparing the scores between Table \ref{tab:3models} and Table \ref{tab:loudnorm_eval}, we confirmed that the performance degradation was greatly alleviated by just the simple loudness normalization of the inputs only at the evaluation stage. Note that the networks were not trained with loudness normalized inputs. This result shows that the input loudness normalization at the evaluation stage can be a quick and easy solution to get robust results from the pre-trained music source separation networks.

Especially, on \textit{drums} of \textit{Demucs v3-B}, it should be noted that the amount of decrease on \textit{median} SDR score between the test using \textit{musdb-hq} and \textit{XL} was sharply reduced from 2.5 dB in Table \ref{tab:3models} to 0.4 dB in Table \ref{tab:demucs_stems}. This implies that the network is overfitted not only to the contents of the signal, but also to the scale or loudness, especially on \textit{drums}. Note that there was no distinction between the given input sources to the networks except the linear gain difference.

Although the performance degradation was reduced by the input loudness normalization, still there exists the performance degradation due to the signal distortion caused by a limiter. Similar to Section \ref{sec:drums_and_vocals}, this result also supports the necessity of the \textit{LimitAug} for robust music source separation.

\subsection{Analysis on various training strategies} \label{sec:various_strategies}
In this section, we trained the \textit{TFC-TDF-U-net} \cite{choi2020investigating} with various training data creating methods; \textit{(1)} linear gain increasing, \textit{(2)} the proposed \textit{LimitAug}, \textit{(3)} input loudness normalization, and  \textit{(4)} input loudness normalization after the \textit{LimitAug}. Of course, the methods \textit{(3)} and \textit{(4)} includes the input loudness normalization at the evaluation stage for the consistency between train and test domains. We compared the results to check which one is the most powerful way for training robust music source separation networks. For the input normalization, we chose the target reference LUFS value as -14.

In Table \ref{tab:training_strategies}, we confirmed that all of the methods were effective for robust music source separation; every methods showed relatively robust performance on \textit{musdb-L} and \textit{XL}, compared to the baseline. Especially, except the method \textit{(1)} targeting LUFS of a normal distribution following statistics of \textit{musdb-XL}, $\mathcal{N}(\mu_{XL},\sigma_{XL}^{2})$, all methods showed greater performance on \textit{musdb-hq} than the baseline. This result implies that these methods are useful not only for the domain shift, but also for the standard benchmark data.

Furthermore, the methods \textit{(3)} and \textit{(4)}, which prevent the domain shift caused by overall amplitude scale mismatch by input loudness normalization, showed slightly better performances than others. In the stem-wise analysis as presented in Table \ref{tab:best_performing}, though we expected that the \textit{LimitAug} would be of help for \textit{vocals} and \textit{drums}, the method \textit{(4)} was better at \textit{vocals} and \textit{other} than  the method \textit{(3)}.

Overall, it seems obvious that considering the music loudness and heavy dynamic range compression from the training stage is beneficial for robust music source separation. For real world applications, it is highly recommended that not just using the single method we experimented, but using various training methods on different stems or using a bag of models trained with various methods.

\section{Conclusions}
In this study, we questioned the domain shift between the research and the real-world data for music source separation, from the viewpoint of music loudness and heavy dynamic range compression.
To answer this, We first built new evaluation datasets, \textit{musdb-L} and \textit{musdb-XL}, which reflect dynamic range compressed music characteristics and heavy loudness. Then, we confirmed the significant performance degradation of state-of-the-art networks on our datasets. To alleviate this, we conducted various experiments on training data construction strategies, including the proposed \textit{LimitAug} method, and confirmed that the methods using the input loudness normalization only or with the \textit{LimitAug} greatly improved the robustness.
We hope that our proposed methods and evaluation datasets could contribute to future music source separation research and application. 

\section{Acknowledgements}
We appreciate Zafar Rafii, the author of \textit{musdb}, for allowing us to reprocess the original musdb data. We thank Antoine Liutkus, also the author of \textit{musdb}, for giving the creative suggestion on the distribution of our proposed datasets. We are grateful to Ben Sangbae Chon, Keunwoo Choi, and Hyeongi Moon from GaudioLab for their fruitful discussions on our proposed methods. Last but not least, we thank Donmoon Lee, Juheon Lee, Jaejun Lee, Junghyun Koo, and Sungho Lee for helpful feedbacks.

\bibliography{ISMIRtemplate}

% Generated by IEEEtran.bst, version: 1.14 (2015/08/26)
\begin{thebibliography}{10}
\providecommand{\url}[1]{#1}
\csname url@samestyle\endcsname
\providecommand{\newblock}{\relax}
\providecommand{\bibinfo}[2]{#2}
\providecommand{\BIBentrySTDinterwordspacing}{\spaceskip=0pt\relax}
\providecommand{\BIBentryALTinterwordstretchfactor}{4}
\providecommand{\BIBentryALTinterwordspacing}{\spaceskip=\fontdimen2\font plus
\BIBentryALTinterwordstretchfactor\fontdimen3\font minus
  \fontdimen4\font\relax}
\providecommand{\BIBforeignlanguage}[2]{{%
\expandafter\ifx\csname l@#1\endcsname\relax
\typeout{** WARNING: IEEEtran.bst: No hyphenation pattern has been}%
\typeout{** loaded for the language `#1'. Using the pattern for}%
\typeout{** the default language instead.}%
\else
\language=\csname l@#1\endcsname
\fi
#2}}
\providecommand{\BIBdecl}{\relax}
\BIBdecl

\bibitem{dredge2013pop}
S.~Dredge, ``Pop music is louder, less acoustic and more energetic than in the
  1950s,'' \emph{The Guardian}, 25 Nov. 2013.

\bibitem{milner2019they}
G.~Milner, ``They really don’t make music like they used to,'' \emph{The New
  York Times}, 7 Feb. 2019.

\bibitem{croghan2012quality}
N.~B. Croghan, K.~H. Arehart, and J.~M. Kates, ``Quality and loudness judgments
  for music subjected to compression limiting,'' \emph{The Journal of the
  Acoustical Society of America}, vol. 132, no.~2, pp. 1177--1188, 2012.

\bibitem{vickers2010loudness}
E.~Vickers, ``The loudness war: Background, speculation, and recommendations,''
  in \emph{Audio Engineering Society Convention 129}.\hskip 1em plus 0.5em
  minus 0.4em\relax Audio Engineering Society, 2010.

\bibitem{stikvoort1986digital}
E.~F. Stikvoort, ``Digital dynamic range compressor for audio,'' \emph{Journal
  of the Audio Engineering Society}, vol.~34, no. 1/2, pp. 3--9, 1986.

\bibitem{stoter19}
\BIBentryALTinterwordspacing
F.-R. St{\"o}ter, S.~Uhlich, A.~Liutkus, and Y.~Mitsufuji, ``Open-unmix - a
  reference implementation for music source separation,'' \emph{Journal of Open
  Source Software}, 2019. [Online]. Available:
  \url{https://doi.org/10.21105/joss.01667}
\BIBentrySTDinterwordspacing

\bibitem{katz2003mastering}
B.~Katz and R.~A. Katz, \emph{Mastering audio: the art and the science}.\hskip
  1em plus 0.5em minus 0.4em\relax Butterworth-Heinemann, 2003.

\bibitem{bregitzer2008secrets}
L.~Bregitzer, \emph{Secrets of recording: Professional tips, tools \&
  techniques}.\hskip 1em plus 0.5em minus 0.4em\relax Routledge, 2008.

\bibitem{uhlich2017improving}
S.~Uhlich, M.~Porcu, F.~Giron, M.~Enenkl, T.~Kemp, N.~Takahashi, and
  Y.~Mitsufuji, ``Improving music source separation based on deep neural
  networks through data augmentation and network blending,'' in \emph{2017 IEEE
  International Conference on Acoustics, Speech and Signal Processing
  (ICASSP)}.\hskip 1em plus 0.5em minus 0.4em\relax IEEE, 2017, pp. 261--265.

\bibitem{pretet2019singing}
L.~Pr{\'e}tet, R.~Hennequin, J.~Royo-Letelier, and A.~Vaglio, ``Singing voice
  separation: A study on training data,'' in \emph{ICASSP 2019-2019 ieee
  international conference on acoustics, speech and signal processing
  (icassp)}.\hskip 1em plus 0.5em minus 0.4em\relax IEEE, 2019, pp. 506--510.

\bibitem{itu2011itu}
R.~ITU-R, ``Itu-r bs. 1770-2, algorithms to measure audio programme loudness
  and true-peak audio level,'' \emph{International Telecommunications Union,
  Geneva}, 2011.

\bibitem{ebu2011loudness}
R.~EBU-Recommendation, ``Loudness normalisation and permitted maximum level of
  audio signals,'' \emph{European Broadcasting Union}, 2011.

\bibitem{musdb18}
\BIBentryALTinterwordspacing
Z.~Rafii, A.~Liutkus, F.-R. St{\"o}ter, S.~I. Mimilakis, and R.~Bittner, ``The
  {MUSDB18} corpus for music separation,'' Dec. 2017. [Online]. Available:
  \url{https://doi.org/10.5281/zenodo.1117372}
\BIBentrySTDinterwordspacing

\bibitem{musdb18-hq}
\BIBentryALTinterwordspacing
Z.~Rafii, A.~Liutkus, F.-R. Stöter, S.~I. Mimilakis, and R.~Bittner,
  ``Musdb18-hq - an uncompressed version of musdb18,'' Aug. 2019. [Online].
  Available: \url{https://doi.org/10.5281/zenodo.3338373}
\BIBentrySTDinterwordspacing

\bibitem{manilow2019cutting}
E.~Manilow, G.~Wichern, P.~Seetharaman, and J.~Le~Roux, ``Cutting music source
  separation some slakh: A dataset to study the impact of training data quality
  and quantity,'' in \emph{2019 IEEE Workshop on Applications of Signal
  Processing to Audio and Acoustics (WASPAA)}.\hskip 1em plus 0.5em minus
  0.4em\relax IEEE, 2019, pp. 45--49.

\bibitem{liutkus20172016}
A.~Liutkus, F.-R. St{\"o}ter, Z.~Rafii, D.~Kitamura, B.~Rivet, N.~Ito, N.~Ono,
  and J.~Fontecave, ``The 2016 signal separation evaluation campaign,'' in
  \emph{International conference on latent variable analysis and signal
  separation}.\hskip 1em plus 0.5em minus 0.4em\relax Springer, 2017, pp.
  323--332.

\bibitem{bittner2014medleydb}
R.~M. Bittner, J.~Salamon, M.~Tierney, M.~Mauch, C.~Cannam, and J.~P. Bello,
  ``Medleydb: A multitrack dataset for annotation-intensive mir research.'' in
  \emph{ISMIR}, vol.~14, 2014, pp. 155--160.

\bibitem{SiSEC18}
F.-R. St{\"o}ter, A.~Liutkus, and N.~Ito, ``The 2018 signal separation
  evaluation campaign,'' in \emph{Latent Variable Analysis and Signal
  Separation: 14th International Conference, LVA/ICA 2018, Surrey, UK}, 2018,
  pp. 293--305.

\bibitem{defossez2021hybrid}
A.~D{\'e}fossez, ``Hybrid spectrogram and waveform source separation,''
  \emph{arXiv preprint arXiv:2111.03600}, 2021.

\bibitem{choi2020investigating}
W.~Choi, M.~Kim, J.~Chung, D.~Lee, and S.~Jung, ``Investigating u-nets with
  various intermediate blocks for spectrogram-based singing voice separation,''
  in \emph{21th International Society for Music Information Retrieval
  Conference, ISMIR, Ed}, 2020.

\bibitem{hennequin2020spleeter}
R.~Hennequin, A.~Khlif, F.~Voituret, and M.~Moussallam, ``Spleeter: a fast and
  efficient music source separation tool with pre-trained models,''
  \emph{Journal of Open Source Software}, vol.~5, no.~50, p. 2154, 2020.

\bibitem{mitsufuji2021music}
Y.~Mitsufuji, G.~Fabbro, S.~Uhlich, and F.-R. St{\"o}ter, ``Music demixing
  challenge 2021,'' \emph{arXiv preprint arXiv:2108.13559}, 2021.

\bibitem{kim2021kuielab}
M.~Kim, W.~Choi, J.~Chung, D.~Lee, and S.~Jung, ``Kuielab-mdx-net: A two-stream
  neural network for music demixing,'' \emph{arXiv preprint arXiv:2111.12203},
  2021.

\bibitem{pedalboard}
\BIBentryALTinterwordspacing
Spotify, ``pedalboard : A python library for manipulating audio.'' 2022.
  [Online]. Available: \url{https://github.com/spotify/pedalboard}
\BIBentrySTDinterwordspacing

\bibitem{steinmetz2021pyloudnorm}
C.~J. Steinmetz and J.~D. Reiss, ``pyloudnorm: {A} simple yet flexible loudness
  meter in python,'' in \emph{150th AES Convention}, 2021.

\bibitem{vincent2006performance}
E.~Vincent, R.~Gribonval, and C.~F{\'e}votte, ``Performance measurement in
  blind audio source separation,'' \emph{IEEE transactions on audio, speech,
  and language processing}, vol.~14, no.~4, pp. 1462--1469, 2006.

\bibitem{le2019sdr}
J.~Le~Roux, S.~Wisdom, H.~Erdogan, and J.~R. Hershey, ``Sdr--half-baked or well
  done?'' in \emph{ICASSP 2019-2019 IEEE International Conference on Acoustics,
  Speech and Signal Processing (ICASSP)}.\hskip 1em plus 0.5em minus
  0.4em\relax IEEE, 2019, pp. 626--630.

\end{thebibliography}

\end{document}